\documentclass[aps,prb,reprint,longbibliography,superscriptaddress]{revtex4-1}

 \usepackage{amsmath,bm}
 \usepackage{mathrsfs}
 \usepackage{amsfonts}
 \usepackage{soul}
 \usepackage{setspace} 
 \usepackage{graphicx}
 \usepackage{epstopdf}
 \usepackage{dcolumn}
 \usepackage{amsmath}
 \usepackage{epsfig}
 \usepackage{indentfirst}
 \usepackage{psfrag}
 \usepackage{subfigure}
 \usepackage{amssymb}
 \usepackage{color}
 \usepackage{units} 
\usepackage{physics}
\usepackage{dcolumn}
\usepackage{bm}
\usepackage{bbold}
 \usepackage{dcolumn}
 \usepackage{natbib}
 \usepackage{wrapfig}
 \usepackage{xcolor}

\usepackage[backref=none,bookmarksnumbered=true,bookmarks=true,bookmarksopen=true,colorlinks=true,citecolor=blue,linkcolor=blue,anchorcolor=green,urlcolor=blue,unicode=false]{hyperref}

\newcommand{\bfp}{\mathbf{p}}
\newcommand{\bfr}{\mathbf{r}}

\usepackage{ulem}[normalem] 

\makeatletter
\newcommand\colorsout[1]{\bgroup \markoverwith{\textcolor{#1}{\rule[0.5ex]{2pt}{0.4pt}}}\ULon}

\makeatother
\begin{document}

\title{Yu-Shiba-Rusinov states in 2D superconductors with arbitrary Fermi contours}

\author{Jon Ortuzar }
     \affiliation{Centro de Física de Materiales (CFM-MPC), Centro Mixto CSIC-UPV/EHU, 20018 San Sebastián, Spain}
     \affiliation{CIC nanoGUNE-BRTA, 20018 Donostia-San Sebasti\'an, Spain}
  
\author{Stefano Trivini}
  \affiliation{CIC nanoGUNE-BRTA, 20018 Donostia-San Sebasti\'an, Spain}  
  
\author{Miguel Alvarado}
\affiliation{Departamento de Física Teórica de la Materia Condensada C-V, Condensed Matter Physics Center (IFIMAC) and Instituto Nicolás Cabrera, Universidad Autónoma de Madrid, E-28049 Madrid, Spain}

\author{Mikel Rouco}
    \affiliation{Centro de Física de Materiales (CFM-MPC), Centro Mixto CSIC-UPV/EHU, 20018 San Sebastián, Spain}
 
\author{Javier Zaldivar}
   \affiliation{CIC nanoGUNE-BRTA, 20018 Donostia-San Sebasti\'an, Spain}

\author{Alfredo Levy Yeyati}
\affiliation{Departamento de Física Teórica de la Materia Condensada C-V, Condensed Matter Physics Center (IFIMAC) and Instituto Nicolás Cabrera, Universidad Autónoma de Madrid, E-28049 Madrid, Spain}

 \author{Jose Ignacio Pascual}
    \affiliation{CIC nanoGUNE-BRTA, 20018 Donostia-San Sebasti\'an, Spain}
        \affiliation{Ikerbasque, Basque Foundation for Science, 48013 Bilbao, Spain}

\author{F. Sebastian Bergeret}
    \affiliation{Centro de Física de Materiales (CFM-MPC), Centro Mixto CSIC-UPV/EHU, 20018 San Sebastián, Spain}
	\affiliation{Donostia International Physics Center (DIPC), 20018 Donostia-San Sebasti\'an, Spain}
\begin{abstract}

Magnetic impurities on a superconductor induce sub-gap Yu-Shiba-Rusinov (YSR) bound states, localized at the impurity site and fading away from it for distances up to several nanometers. In this article, we present a  theoretical method to calculate the spatial distribution of the YSR spectrum of a two-dimensional superconductor with arbitrary Fermi contours (FCs) in the presence of magnetic impurities. Based on the  Green’s Function (GF) formalism, we obtain a general analytical expression by approximating  an  arbitrary contour shape to a regular polygon. This method allows us to show the connection between the spatial decay (and, hence, the extension) of YSR  states  and the shape of the FC of the host superconductor.  We demonstrate the accuracy of this approximation by comparing the results with those obtained from an exact numerical calculation based on a tight-binding Hamiltonian. 
We further apply the analytical formalism to compute the evolution of YSR states in the presence of a nearby impurity atom, and compare the results with Scanning Tunneling Microscopy (STM) measurements on interacting manganese dimers  on the $\beta-$Bi$_2$Pd superconductor.  The method can be easily extended to any arbitrary number of magnetically coupled impurities, thus providing a useful  tool for simulating the spectral properties of interacting YSR states in artificial atomic nanostructures. 
\end{abstract}
\date{\today}
\maketitle

\section{Introduction}
Magnetic impurities placed on a superconductor are a potential workbench for investigating many fundamental properties of the superconductor and  pairing mechanisms. As predicted a long time ago independently by Yu, Shiba, and Rusinov~\cite{Yu,Shiba,Rusinov}, 
the pair-breaking potential induced by a magnetic moment distorts the superconducting bands and gives rise to superconducting bound states with sub-gap quasiparticle excitations, so-called YSR states. Because they lie within the band gap, these excitations are long lived and, hence, appear as very narrow resonances in tunneling experiments~\cite{Ji2008,Heinrich2018}. YSR states are the elementary states forming  subgap bands in atomic chains of magnetic impurities  \cite{Nadj-Perge2014,kim2014helical,Ruby2015a,peng2015strong,Pawlak2016,Ruby2017,Schneider2021c,Mier2021,Schneider2021d,liebhaber2021}, allegedly turning topological and hosting  Majorana  bound states at their ends. The  study of YSR states also reveals fundamental aspects of atomic scale magnetism.  For example, STM experiments on superconductors with magnetic impurities revealed valuable information about molecular \cite{Franke2011,Hatter2015,Hatter2017,Farinacci2018, Kezilebieke2018,Brand2018,Kezilebieke2019,Malavolti2018,Farinacci2020,Rubio-Verdu2021,Song2021a}  and atomic properties, as hybridization of orbitals, anisotropy, etc.\cite{Cornils2017,Kamlapure2018,Senkpiel2019a,Song2020a,Yang2020a,Liebhaber2019,Huang2020,Huang2020c,Odobesko2020,Ding2021,Beck2021,Chatzopoulos2021,Huang2021,Wang2021,Friedrich2021a,Kamlapure2021,Kuster2021}.

The spatial distribution of YSR states are the subject of several recent studies based on STM. YSR  states appear localized around atomic-sized magnetic impurities on superconductors, and their spatial distribution may reflect the shape of atomic or molecular orbitals responsible of the YSR channel \cite{Ruby2016,Choi2017,Farinacci2020,Rubio-Verdu2021,Beck2021,Schneider2021c}. Furthermore, the YSR states are generated by the local exchange interaction between the impurity orbital with  electrons in the superconducting host  and, consequently, far from the impurity the distribution of YSR amplitude reflects the shape of superconducting bands\cite{Menard2015,Ruby2016,Kezilebieke2018,Etzkorn2018a,Liebhaber2019,Kim2020}. 

As demonstrated by Rusinov in his original work \cite{Rusinov}, there are two length scales involved in the spatial decay of the bound states away from the impurity: a long-range scale determined by the superconducting coherence length $\xi_s$, which provides an exponential decay, $e^{-x/\xi_S}$ of the YSR wavefunction, and a short length equal to the inverse of the Fermi momentum $k_F^{-1}$, over which the wave function oscillates and exhibits an algebraic decay $\sim (k_F r)^{-1}$. This  decay law is valid for three-dimensional systems. For a general isotropic superconductor, theory predicted that the decay law depends on dimensionality of the superconducting band, namely the YSR amplitude decays as $\sim (k_F r)^{(1-d)/2}$ with dimension $d=1,2,3$  \cite{Menard2015}. 

Several STM experiments observed YSR decaying several nanometers in conventional superconducting systems \cite{Menard2015,Ruby2016,Etzkorn2018a,Kim2020}, which were attributed to a combination of two effects, a reduced dimensionality of the superconducting bands and the anisotropy character of their Fermi surfaces or contours. This last effect is connected with the accumulation of multiple scattering wavevectors along specific directions of the substrate, due to the existence of flat segments in the Fermi surface. These  cause an "electronic focusing" effect along specific directions \cite{Weismann2009}, resulting in the propagation of the wavefunction amplitude for larger distances. In spite of the clear relevance of this effect, specially for highly anisotropic superconductors, there is not a detailed analytical study on its role in the decay of superconducting quasiparticle (QP) states. To incorporate complex Fermi surfaces and contours one relies on  high-throughput numerical simulations.

In this article we present an analytical model based on   the well-established GF technique to study the spatial dependence of the YSR spectrum in superconductors with  a non-spherical FC and several impurities. Specifically, we focus on two-dimensional superconductors with a FC that can be approximated by a regular polygon. In the limit of the Fermi energy being much larger than the superconducting gap we obtain analytical expressions for the GF of a superconductor with arbitrary FC, which can be approximated by a N-sided regular polygon, from which one can  determine the local YSR spectrum of magnetic impurities inside.
We use our result to obtain explicitly  the GF's of a  square and hexagonal shaped FC \cite{ZaldivarPhD,Kim2020}. These cases are good approximations to describe surface superconductivity in  materials like  $\beta$-Bi$_2$Pd and NbSe$_2$ respectively. 

In section \ref{sec:DIM}, we use our analytical  method  to study the space dependence of the YSR amplitude away from the scattering impurity. The mentioned \textit{wave-vector focusing} effect shows up in experiments as 
an anisotropic standing wave pattern of YSR amplitude decaying away from the magnetic impurity.  We reproduce YSR wavefunction oscillations along specific directions of the surface depending on the shape of the FC. In particular, we confirm  the larger extension or smaller distance-decay of YSR oscillation pattern scales with the size of flat segments in the FC, i.e. with the density of nesting vectors along one dimension. In section \ref{YSR_TB}, we compare the analytical method to a tight-binding superconducting model in a square lattice which, depending on the chemical potential, can hosts circular or squared FCs.

Finally, in section \ref{2YSR} we apply our model to study the hybridization of YSR states from several impurities. We find that interacting YSR states interfere and split in odd and even states, following predictions by numerical methods \cite{Flatte2000,Kezilebieke2018}. We further  calculate the dependence of the YSR energy splitting with the relative distance and angle between adatoms. This simple simulations can be utilized to extract information about the type of magnetic alignment between spins. To illustrate this, we compare the theoretical results with experiments on Mn dimers on $\beta$-Bi$_2$Pd and obtain information about the type of magnetic coupling,  ferromagnetic (FM) or anti-ferromagnetic (AF), of  the dimers in different surface configurations.

\section{The Model} 
\label{sec:model}
In this section we  use  the well-established GF method to  build up a general expression for the density of states of a two dimensional superconductor with an arbitrary FC shape, modeled as a polygon. We also discuss how to treat multiple atomic impurities by deriving a compact expression for the GF in real space.

\subsection{Local density of states}

We consider a two-dimensional BCS superconductor described by the Bogoliubov-de Gennes Hamiltonian\cite{de1966superconductivity}
\begin{equation}
    \check{\mathcal H}_0(\bfr) = \xi(\hat\bfp) \hat\tau_3 + \Delta \hat\tau_1\; ,
    \label{eq:free-hamiltonian}
\end{equation}
where $\xi(\bfr)$ is the  quasiparticle's energy operator, $\Delta$ is the s-wave superconducting gap, and  $\hat\bfp = -i\hbar\nabla_\bfr$ is the momentum operator.   Hamiltonian \eqref{eq:free-hamiltonian} is a 4$\times$4 matrix in the Nambu$\times$Spin space. In the absence of impurities it has a trivial structure in spin space, whereas the Nambu structure is described by the Pauli $\hat \tau_i$  matrices. In our notation  2$\times$2 and 4$\times$4 matrices are indicated with ``hat'' ($\hat\cdot$) and ``check'' ($\check \cdot$) symbols, respectively.

We introduce   \textit{N} point-like  magnetic impurities,  
located at $\bfr_n$ , where  the index  $n$ 
is  the impurity index. The potential of each impurity is described by 
\begin{equation}
    \check V_n = U_n \hat\tau_3 
    + h_n^a\hat\sigma^a\; ,
    \label{eq:impurity-potential}
\end{equation}
where $\hat\sigma^a$ stand for the Pauli matrices spanning the spin space, and $U_n$ and $h_n^a$ stand for the electrostatic and a-spin component of the exchange fields in the $n$-th magnetic impurity, respectively. 

In order to obtain the spectrum of the system we introduce  the equation of motion for the 4$\times$4 matrix GF, the so called  Gor'kov equation\cite{abrikosov2012methods} which, in real frequency $\epsilon$ space,  reads,
\begin{equation}
    \Big[\epsilon - \check{\mathcal H}_0(\bfr) - \sum_{n=1}^N \check V_n \delta(\bfr - \bfr_n) \Big] \; 
    \check G(\bfr, \bfr';\epsilon) = \delta(\bfr - \bfr')\; . 
    \label{eq:gorkov}
\end{equation}
We obtain the retarded and advanced GF by adding an infinitesimal  $\eta$ to the frequency, $\epsilon\rightarrow\epsilon\pm i \eta$, respectively.   Once the GF is known  one can compute  the local density of states ($\rho(\bfr, \epsilon)$) from the retarded GF: 
\begin{equation}
\label{eq:ldos}
    \rho(\bfr, \epsilon) = \frac{1}{4\pi} \text{Tr} \Big[ 
    \text{Im} \check G(\bfr, \bfr,\epsilon+i\eta) \Big]\; ,
\end{equation}
where  the trace runs over the Nambu$\times$spin space.

\subsection{General expression for the GF in the presence of multiple impurities}\label{sec:mult_imp}

The spectrum of a single impurity can be found explicitly  by solving Eq. (\ref{eq:gorkov}) analytically. In the case of multiple impurities the equations become more  cumbersome. In this subsection we provide a useful expression to compute $\check G(\bfr, \bfr)$ in presence of $N$ impurities situated at arbitrary position ${\bf r}_i$.

We start writing the solution of   Eq.~\eqref{eq:gorkov} in the  form of a Dyson series:
\begin{equation}
    \check G(\bfr, \bfr') = \check G_0(\bfr - \bfr') 
    + \sum_n \check G_0(\bfr - \bfr_n) \check V_n 
    \check G(\bfr_n,\bfr'),
    \label{eq:dyson-equation}
\end{equation}
where $\check G_0(\bfr-\bfr')$ is the GF of the 2D superconductor without impurities. To simplify the  notation we drop the $\epsilon$-dependence of the GFs. Our goal is to write the right-hand side of this equation only in terms of the unperturbed $\check G_0$.
For this we write  the $N$ equations for the matrices $\check G(\bfr_n, \bfr')$, with $n=1,\dots,N$ in a compact form: 
   \begin{equation}
    \check{\overline{G}}(\bfr') = 
    [\textbf{I} - \textbf{M}]^{-1} \check{\overline{G}}_0(\bfr')\; , 
    \label{eq:shorthand-big-system}
\end{equation}
where is $\textbf{I}$ is the $4N \times 4N$ identity matrix and  we have introduced the shorthand notation:
\begin{widetext}
\begin{equation}
  \check{\overline{G}}(\bfr') = 
  \left(\begin{array}{c}
      \check G(\bfr_1, \bfr') \\[.3em]
      \check G(\bfr_2, \bfr') \\
      \vdots \\
      \check G(\bfr_N, \bfr') \\[.3em]
  \end{array}\right), 
  \quad
  \check{\overline{G}}_0(\bfr') = 
  \left(\begin{array}{c}
      \check G_0(\bfr_1 - \bfr') \\[.3em]
      \check G_0(\bfr_2 - \bfr') \\
      \vdots \\
      \check G_0(\bfr_N - \bfr') \\[.3em]
  \end{array}\right),
  \end{equation}
  and the $4N \times 4N$ matrix
  \begin{equation}
  \textbf{M} = \left(
  \begin{array}{cccc}
    \check G_0(0) \check V_1 & \check G_0(\bfr_1 - \bfr_2) \check V_2 & \cdots & \check G_0(\bfr_1 - \bfr_N) \check V_N \\[.3em]
    \check G_0(\bfr_2 - \bfr_1) \check V_1 & \check G_0(0) \check V_2 & \cdots & \check G_0(\bfr_2 - \bfr_N) \check V_N \\
    \vdots & \vdots & \ddots & \vdots \\
    \check G_0(\bfr_N - \bfr_1) \check V_1 & \check G_0(\bfr_N - \bfr_2) \check V_2 & \cdots & \check G_0(0) \check V_N \\[.3em]
  \end{array}\right).
  \label{eq:big-matrices}
\end{equation}
\end{widetext}
The matrix $(\textbf{I} - \textbf{M})$ contains  the information about the bound states of a system of N magnetic adatoms and their hybridization. In particular the YSR states are determined from the condition $\text{det}(\textbf{I} - \textbf{M}) = 0$. 

By solving the set of equations \eqref{eq:shorthand-big-system},  and after substitution into Eq.~\eqref{eq:dyson-equation} one can obtain the full GF in terms of $G_0$, the GF in the absence of impurities.  In most of previous works  superconductors with a spherical Fermi surface were considered.  In the next subsection we obtain $\check G_0(\bfr - \bfr')$ for FCs with arbitrary shape that 
can be approximated by a polygon. 
Moreover, in section \ref{2YSR} we use   Eqs. (\ref{eq:dyson-equation}-\ref{eq:big-matrices}) for the two impurities case ($N=2$).

\subsection{ Real space GF for  superconductors with non-circular FC} 
\label{FC}
In this section we obtain the GF of a 2D superconductor with an arbitrary FC. We  calculate $\check G_0(\bfr - \bfr')$ in a $M$-sided regular polygon centered at $\bfp = 0$ (Appendix~\ref{app:A}). We focus on three particular examples -- the square ($M=4$), hexagon ($M=6$), and circular ($M \rightarrow \infty$) FCs -- and compare the spatial decay of the states bounded to the impurities in  all cases.

\begin{figure}[b!]
\centering
	\includegraphics[width=1\columnwidth]{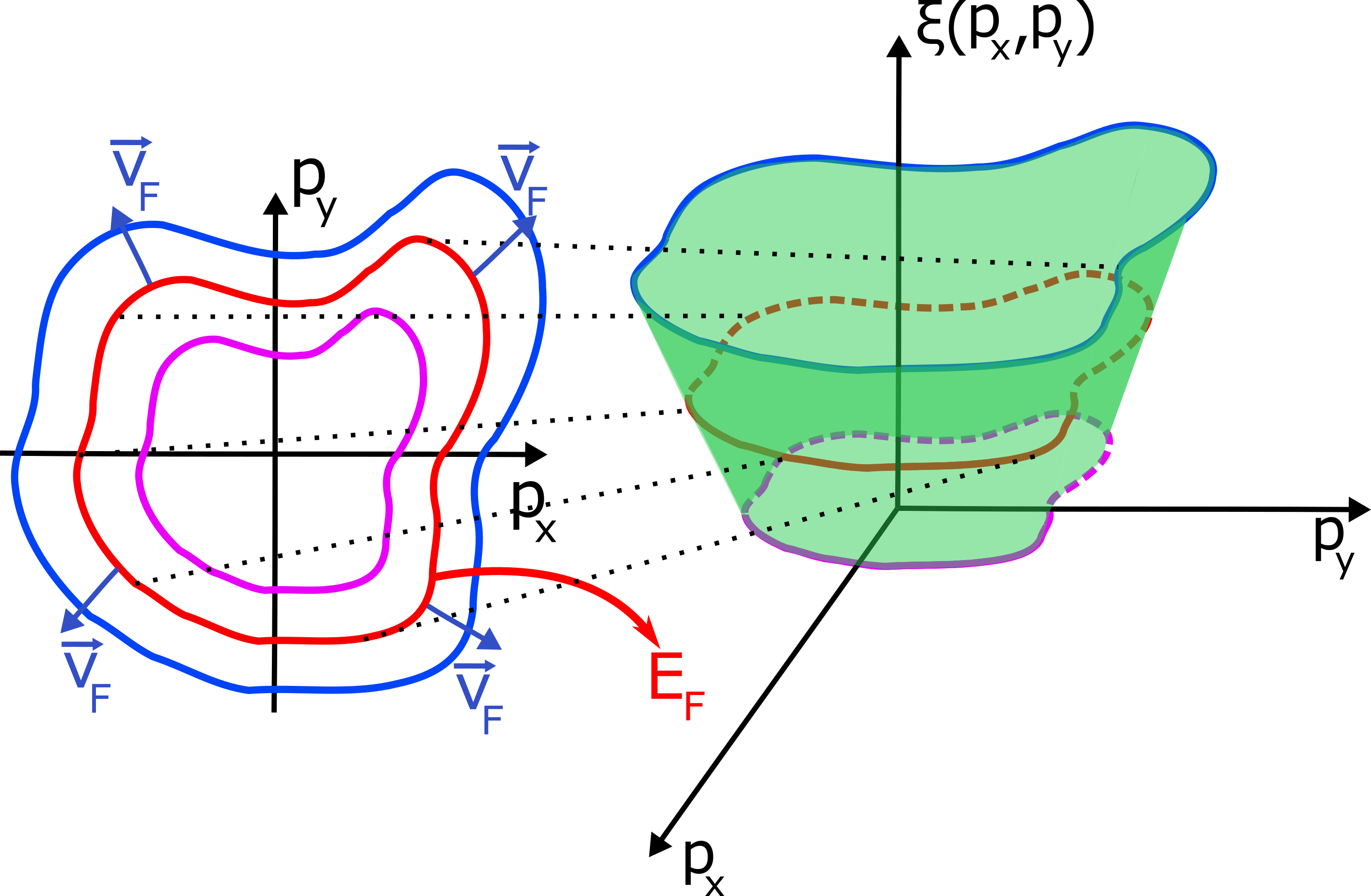}
	\caption{Right panel: Sketch of the quasiparticle energy versus two dimensional momentum {\bf p}. The red curve, at $\xi=0$, is the FC. Left panel:  sketch of the FC on the $(p_x,p_y)$ plane. The vector $\textbf{v}_F$ is parallel  to  $\boldsymbol{\nabla}_{\textbf{p}}\xi$ and hence points in the direction perpendicular to the curves of equal energy.  }
	\label{Fig0} 
\end{figure}

The real space GF of the two dimensional clean superconductor reads,
\begin{equation}\label{fouriertrans}
    \check G_0(\bfr - \bfr')=\int\dfrac{d^2\textbf{p}}{(2\pi)^2} 
    \check G_0(\bfp) e^{i\bfp \cdot (\bfr - \bfr')},
\end{equation}
where $\check G_0(\bfp)$ is given by the Gor'kov equations of the hosting media in momentum space, $[\epsilon - \xi(\bfp)\hat\tau_3 - \Delta\hat\tau_1] \check G_0(\bfp) = 1$. We transform the integral in Eq. (\ref{fouriertrans}) to an integral over the quasiparticle energy $\xi$, by   writing  $d^2p= d{\cal C} dp_n$, where $d{\cal C} $ is a  differential element on a constant energy contour and $p_n$  the perpendicular component of the momentum, normal to such contour, with $dp_n=d\xi/|\partial \xi/\partial {\bf p}|$,  see sketch in Fig. \ref{Fig0}.
The relevant contribution to the integral is around the FC, where the GFs have  poles. Therefore it is convenient to linearize $\xi$ around the FC. 
\begin{equation}
    \xi(\bfp) \approx 
    \textbf{v}_F \cdot (\bfp-\bfp_F),
\end{equation}
where  the Fermi velocity, $\textbf{v}_F \equiv \nabla_\bfp \xi\big|_{\bfp_F}$, points in the direction perpendicular to the the constant energy contours (see sketch in Fig. \ref{Fig0}). The  integral over $\xi$ goes from $-\mu$ to $\infty$. In metallic systems, $\mu$ is usually the largest energy scale, hence we take the limit $\mu\rightarrow\infty$ and integrate using residue theorem.  

In the next section we compute  the  integral \eqref{fouriertrans} for inscribed regular polygons, and discuss some examples.

\begin{figure*}
    \begin{center}
	\includegraphics[width=\textwidth]{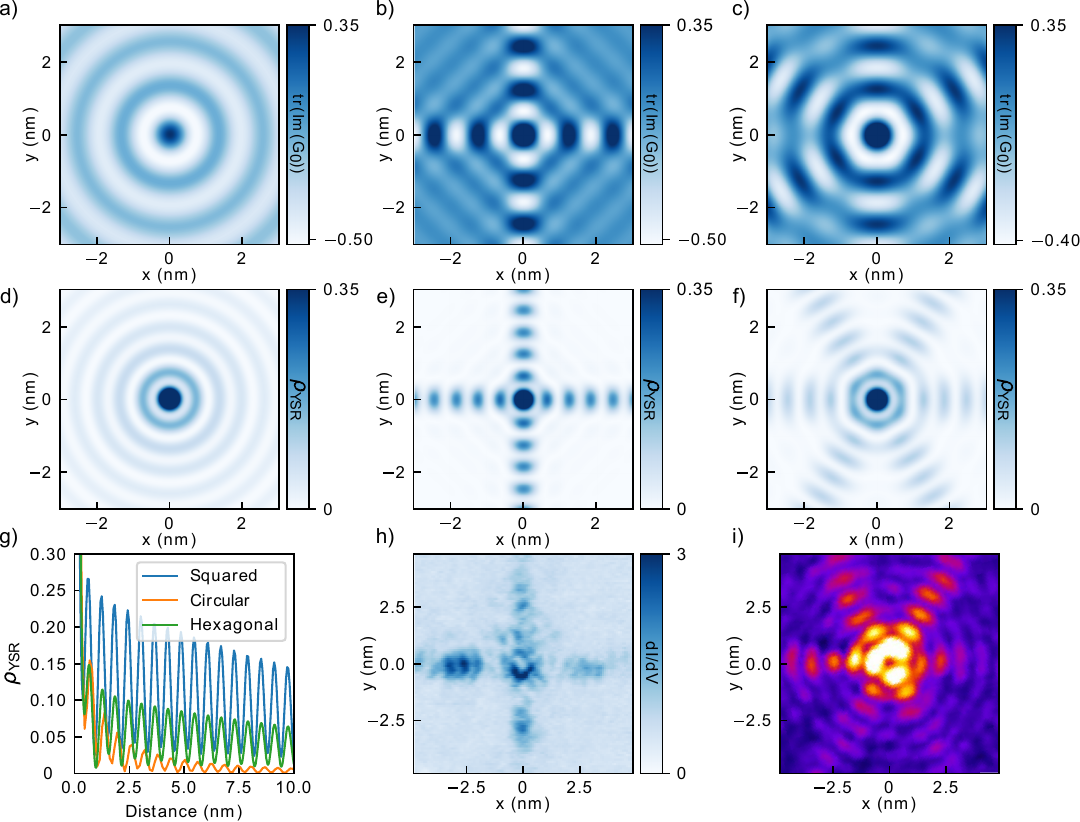}
	\end{center}
	\caption{(a)-(c)  The spatial dependence of the correlation function  $Tr[Im[G_0(x,y;\epsilon+i\eta)]]$ evaluated at $\epsilon=2\Delta$ and normalized with respect to its value at $(x,y)=(0,0)$: (a) for a circular , (b) for a square , and (c) for a hexagonal FC. (d)-(f) YSR calculated DoS for the three different FCs and normalized with respect to its value at  $x=y=0$. (e) Radial cut along the (100) direction of the $\rho_{\rm{YSR}}$ in panels (d)-(f). (h) Conductance map recorded by STS for one YSR state of a isolated V adatom deposited on $\beta-$Bi$_2$Pd \cite{ZaldivarPhD}. Parameters: I=250 pA, V=0.93 mV. (i) YSR spatial dependence measured  for magnetic impurities  on La(1000) films grown on a Re(1000) crystal [adapted from figure 1a in Ref. \onlinecite{Kim2020}]. }
	\label{Fig1} 
\end{figure*}

\section{Effect of the dimensionality and shape of  FC  on  the  YSR states spatial dependence}\label{sec:DIM}

The dimension of the host superconductor determines  the characteristic of the decay of the superconducting GF \cite{Menard2015}. The latter  manifests on the spatial dependence of the YSR peaks.  
Namely,  the spatial dependence of the normal component of the GF is a fast oscillating function with period $k_Fr $ with an exponential decay   over  the coherence length $\xi_s$. For an  spherical  FC, there is an additional pre-factor  $(k_Fr)^{(1-d)/2}$, where $d$ is the dimensionality of the system. From which it follows that, lower dimensional systems exhibits longer correlation distances. 

According to this rule, in a 2D superconductor with a circular FC the decay of the GF obeys $(k_Fr)^{-1/2}e^{-r/\xi_s}$, see   Eq. \eqref{esph_asymp} of the Appendix. If we instead assume that the FC can be approximated by an M-sided regular polygon, we can determine the GF in the absence of impurities following the procedure described in section \ref{FC}. For the unperturbed GF we obtain the following  expression ({\it cf.} Eq. (\eqref{cg0})) 
\begin{widetext}
\begin{equation}\label{G0-Msided}
    G_0(x,y)=\sum_{k=1}^{M}\dfrac{1}{(2\pi)^2}\int_{n^k/w^k-\tan\frac{\pi}{M}}^{n^k/w^k+\tan\frac{\pi}{M}}d\chi \int d\xi  G_0(\xi)e^{-i(\frac{m\xi}{p_F'}+p_F')\chi w^k}\; ,
\end{equation}
\end{widetext}
where  $n^k=x\cos(\frac{2\pi k}{M})+y\sin(\frac{2\pi k}{M})$,  $w^k=-x\sin(\frac{2\pi k}{M})+y\cos(\frac{2\pi k}{M})$,  and $p_F'=p_F\cos(\frac{\pi}{M})$, with $k=1,2,...,M$. It is straightforaward to check, see the Appendix, that in the limit $M\rightarrow\infty$ we recover the result for the circular FC. 

We first use Eq. (\ref{G0-Msided}) to calculate the GF for a  square-like FC. The exact expressions is given in the Appendix (eq. \ref{G0-square}). To illustrate the anisotropic spatial behaviour of the GF we compute $G_0$ in two directions: $y=0$ (eq. \eqref{y=0}) and $x=y$ (eq. \eqref{x=y}). These equations show a decaying behaviours that differ from an istropic  2D superconductor. In the diagonal direction, the decay resembles a 3D  superconductor, while, in the $y=0$ direction, one obtains a 1D-like decay. 

The full DoS, including the   
YSR bound states, can be written in terms of the unperturbed Green's functions, $G_0$. For the case of a single YSR state the corresponding GF reads \cite{Balatsky}
\begin{equation}
\label{eq:dysonYSR}
    \check G(\textbf{r},\textbf{r}')=\check G_0(\textbf{r}-\textbf{r}')+\check G_0(\textbf{r})\check V[\mathbb{1}- \check G_0(0)\check V]^{-1}\check G_0(\textbf{r}')\; .
\end{equation}
The poles of the second term in the right-hand side determines the energy of the YSR bound states, all the  spatial information is contained in the unperturbed GF, $\check G_0 (\bf r)$.

In Figs \ref{Fig1} (a), (b) and (c) we show  the correlation function, ${\rm Tr}[\rm Im\check G_0({\bf r})]$ at an energy $\epsilon=2\Delta$, for circular, square-shaped and hexagon-shaped FCs, respectively. The square- and hexagon-shaped FCs resemble the Fermi surface of the $\beta$-Bi$_2$Pd\cite{Sakano2015} and  La(0001)\cite{Kim2020} superconducting surface. We approximate both cases by a single, square- or hexagon-shaped band.  To compare the results, throughout the work we use the same set of parameters. Namely, for the superconducting gap $\Delta = 0.78 mV$, effective mass $m=6.67m_e$, Fermi momentum $p_f=0.274/a_0$ ($a_0=3.3\dot{A}$), and exchange coupling $\alpha^2=(\pi \nu_0JS)^2= 0.156$.

We next focus on the spatial distribution of the YSR states for the same three  examples. In  Figs. \ref{Fig1} (d), (e) and (f),
we show the spatial dependence of the DoS obtained from the full GF, Eqs. (\ref{eq:ldos},\ref{eq:dysonYSR}), evaluated at the energy of the YSR bound state. 
In panel Fig. \ref{Fig1} (g) we show cuts of the DoS along the (100) direction. As expected, the  spatial decay of the circular FC is faster than the one along the symmetrical direction of the square-shaped FC, which behaves as a lower dimensional case.  The hexagon-shaped case lies  in between.  

Finally, in  Figs. \ref{Fig1} (h) and (i) we show STM  measurements. Fig. \ref{Fig1} (h) shows the spatial dependence  of a YSR state created by a V adatom on the surface of $\beta-$Bi$_2$Pd. The latter is a multiband type-II superconductor, with presumably a surface 2D superconductivity. First-principle calculations \cite{iwaya2017full}
suggest  that the  $\beta$-Bi$_2$Pd has square-shaped bands. Qualitatively this is confirmed by comparing Fig. \ref{Fig1} (h) and our predictions  for a (single) square-shaped  FC, Fig. \ref{Fig1} (e). 
In Fig. \ref{Fig1} (h), we show an example of an hexagon-shaped FC. Namely the spatial dependence of the YSR of magnetic impurities in the La(1000) films grown on Re(1000)\cite{Kim2020}. 

\begin{figure}[h]
    \begin{center}
	\includegraphics[width=1\columnwidth]{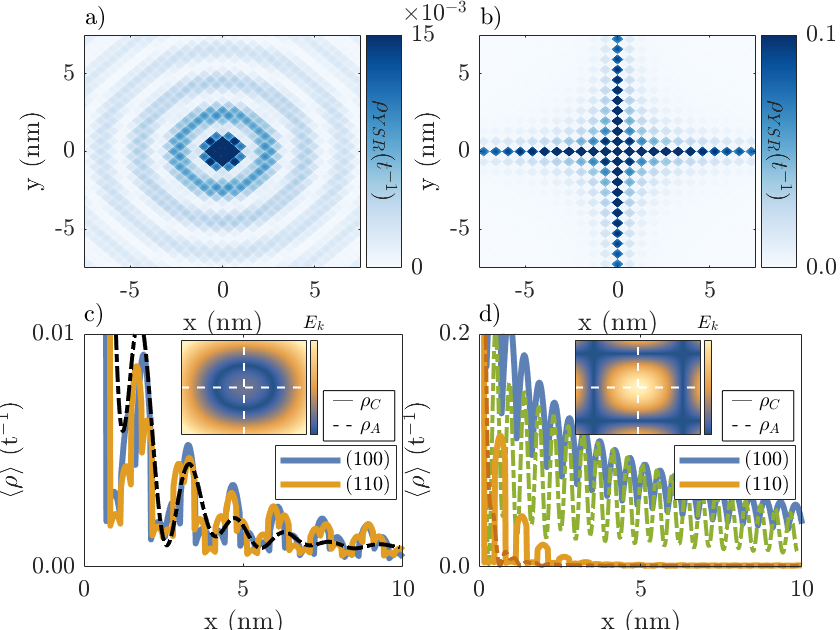}
	\end{center}
	\caption{(a)-(b) The spatial dependence of the DoS  evaluated at the YSR state in units of the inverse of the hopping term: a) for a circular FC with $\mu = 0.5t$, $k_F = 0.71 \pi/a_0$ and $\epsilon=0.72\Delta$, b) for a square FC with $\mu = 4t$, $k_F = 0.22 \pi/a_0$ and $\epsilon=0.82\Delta$. (c)-(d) Radial cut along the (100) and (110) direction of the $\rho_{\rm{YSR}}$ in panels (a)-(b). Solid lines represents the DoS obtained from the computational calculation from the  tight-binding Hamiltonian ($\rho_{C}$). Dot-dashed lines represent the analytical calculation ($\rho_{A}$). (c)-(d) Insets show the band dispersion of the model  for the circular and square-shape FCs. Blue (yellow) in the color-bar represents the minimum (maximum) of the band dispersion. The discrete DoS of the TB calculation is interpolated using a Gaussian distribution centered on the lattice sites. To compare the spatial decay of the YSR states we normalize  the analytical DoS  with respect the DoS around $x=4nm$ in the lattice model.}
	\label{Fig_TB} 
\end{figure}

The good agreement between theoretical (Figs. \ref{Fig1} (e-f)) and experimental (Figs. \ref{Fig1} (h-i)) results demonstrates the suitability of our model for a qualitative description of the $\rho(\bfr, \epsilon)$ of superconductors with magnetic impurities. The results demonstrate  how  the low-symmetric square- and hexagon-shaped FCs lead to a slower decay of the YSR states. Such a decay is similar to the 1D situation, and suggests the use of superconductors with a square-shaped FC for the realization of one-dimensional Andreev crystals \cite{rouco2021gap,rouco2021_PhysRevB.104.064506} by placing chains of magnetic defects along the direction parallel to the symmetry axis of the square. 

\section{Comparison with Tight Binding models} \label{YSR_TB}

 To compare the above analytical results in which FCs are approximated by a regular polygon, we present in this section exact solutions 
of a  superconducting tight-binding model in a square lattice.   By changing  the chemical potential $\mu$ the FC evolves from a circular to a square shape in a continuous manner. The tight-binding Hamiltonian reads
\begin{equation}
    \check{\mathcal H}_{TB}({\bf r}) = \xi({\bf k}) \hat\tau_3 + \Delta \hat\tau_1\; ,
    \label{eq:TB-hamiltonian}
\end{equation}
where $\xi({\bf k}) = -2t(\cos{k_x a_0}+\cos{k_y a_0}-2)-\mu$.  We can rotate the momentum direction or the spatial coordinates in order to obtain a proper squared shape parallel to the coordinate axes. Based on the set of parameters of the continuum models we estimate the hopping term as $t = \hbar^2/(2 m a_0^2) \approx 50mV$.


The local unperturbed GF is obtained from Eq.~\eqref{fouriertrans} now defined in the first Brillouin zone. The integration on one of the momenta is converted into a contour integral and performed using the algorithm described in \cite{alvarado20212d}. As a result we
obtain a regularized GF that can be finally integrated computationally along the other momentum to get the local unperturbed GF.

In Fig.~\ref{Fig_TB} we show  the spatial variation of the local DoS around the impurity evaluated at the  YSR bound state energy $\epsilon\approx0.8\Delta$, for different values of the chemical potential reproducing the limits in which the FCs are compatible with a circular ($\mu = 0.5t$) and square-shape ($\mu = 4t$). Notice that in both cases  $\mu \gg \Delta$ which is the limit at which the analytical approach of Section \ref{sec:DIM} is valid. 
 The Fermi momenta are fixed by the dispersion relation for a given chemical potential and consequently the coherence length for each case is different.  We find a very good quantitative agreement between numerical an analytical approaches which shows that the regular shape approximation can be safely used even when the FCs are not perfect polygons. Furthermore, the analytical approach provides the correct spatial  decay of the YSR  away from the impurity in both circular and square-shaped FCs in $(100)$ and $(110)$ directions.

\section{Hybridization of YSR bound states from neighbouring impurities} \label{2YSR}

Atomic manipulation using the tip of a STM has demonstrated a large potential for fabricating atomic nanostructures of magnetic impurities on superconductor and exploring the hybridization of their YSR wavefunction \cite{Kezilebieke2018,Ruby2018,Choi2018,Kim2018,Ding2021,Beck2021,Kamlapure2021,Kuster2021}. As predicted by Flatté and Reynolds \cite{Flatte2000}, the hybridization between overlapping YSR states depends on the relative alignment of the impurities' spins and, for the case of parallel aligned spins, leads to a splitting of the  sub-gap features into two new states with symmetric and antisymmetric spatial distribution. The splitting oscillates with the  separation between impurities ($d$) with periods comparable with the  Fermi wavelength of the substrate. In the presence of non-isotropic FCs the spatial distribution of the YSR splitting shows intriguing orientation dependence for short inter-impurity distances \cite{Flatte2000,Kezilebieke2018}.

\begin{figure}[h]
	\includegraphics[width=1\columnwidth]{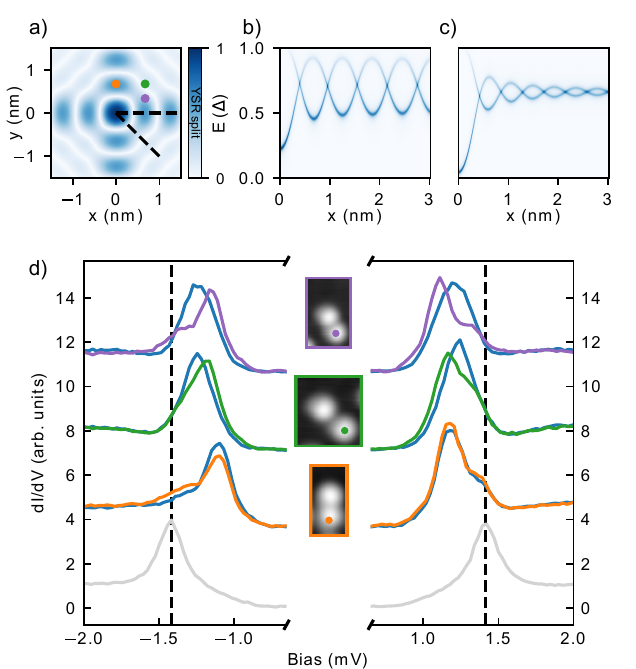}
	\caption{(a)  The YSR splitting energy $E(\Delta)$,  on a S=1/2 impurity as a function of the $x-$ and $y-$position of a second impurity on a 2D  superconductor with a square-shaped FC. The anisotropy of the band determines certain directions along which the splitting is larger.
	(b,c) YSR subgap spectra on the impurity as a function of the position of the second impurity along the (1,0) and (1,-1) lines of high symmetry (indicated with dashed lines in (a)). 
	(d) dI/dV STS measurements with a superconducting tip on a Mn adatom on $\beta$-Bi$_2$Pd before (blue) and after (coloured) formation of a Mn$_2$ dimer by bringing a second Mn adatom into three different substrate positions, with respect to the probed adatom.   From top to bottom: (2$a$,$a$), (2$a$,2$a$) and (0,2$a$), with $a$ being the lattice constant of $\beta$-Bi$_2$Pd (i.e. $d=\sqrt{5}a$, $2\sqrt{2}a$, and $2a$, respectively). These atomic sites are represented in panel (a) as coloured dots. The bottom gray spectra is the reference spectra measured on a bare substrate region (the gap of the $\beta$-Bi$_2$Pd substrate is 0.75 meV and the tip's gap approaches close to this value). }
	\label{Fig2}
\end{figure}
 
In this section we explore the role of the YSR-focusing effect described above  on  the YSR hybridization. We apply the model described in section \ref{sec:mult_imp} to study YSR hybridization of two classical impurities on a surface with a squared FC as a function of their alignment on the substrate. We first assume that the impurities have parallel spin and analyze the splitting of their YSR states as a function of their alignment. The GF of the dimer can be obtained by constructing the  matrix in  Eq.  \ref{eq:big-matrices} with N=2. The resulting spectral function shows that the YSR states split by an amount $E$ representing the hybridization between YSR states. As we show in Fig.~\ref{Fig2}(a) the splitting energy oscillates as a function of the distance between impurities.   In Fig.~\ref{Fig2}(b,c), we show that the oscillation amplitude of the YSR splitting is barely constant when the impurities are aligned along the (100) direction (panel b), while quickly decays along the (110) direction. This proves that the focusing effect enhances the hybridization when the atoms are aligned parallel to the direction of the nesting vectors.  

To correlate these simulations with real systems, we compare them with experimental results on pairs of manganese atoms positioned with precision by means of atomic manipulation on the $\beta$-Bi$_2$Pd superconductor  surface. The $\beta$-Bi$_2$Pd surface is a squared lattice of Bismuth atoms with lattice parameter $a=3,36$~\AA. Adatoms in neighbour sites frequently collapse in Mn$_2$ dimers with no subgap features \cite{Choi2018}. Therefore, we explore the possible next-neighbour distances, namely Mn dimers aligned along the (210), (110) and (100) crystallographic directions with spatial separation $\sqrt{5}a$, $2\sqrt{2}a$ and $2a$. In Fig.~\ref{Fig2}(d) we compare differential conductance spectra measured  on a reference adatom before (blue) and after placing a second adatom at the  position indicated in the insets. When the second adatom is located at the sites (2$a$,$a$) or (2$a$,2$a$) the YSR state appears splitted by $\sim$300~$\mu$eV and $\sim$200~$\mu$eV, respectively. The larger splitting for the former, as well  as the range of the splitting energy, are qualitatively reproduced by the theory, thus suggesting that these dimers have their spin with a  close-to-parallel alignment. 

For the third dimer, the YSR peaks are barely affected by the addition of a Mn adatom on the (0,2$a$) site, while our model in Fig.~\ref{Fig2}(a) predicts a larger splitting than in the previous cases due to the focusing effect.  This suggests that in this configuration the Mn dimers are antiferromagnetically (AF) aligned. As observed in our previous results on this surface \cite{Choi2018}, the substrate-mediated exchange coupling between adatoms also depends on their relative orientation, with a preference of AF alignment along the high-symmetry (100) direction.

\begin{figure}[h]
\includegraphics[width=1\columnwidth]{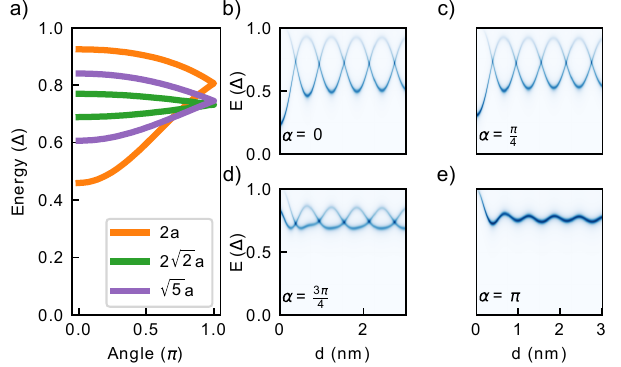}
	\caption{(a) Calculated dependence of the  YSR splitting on the mutual spin orientation,  from 0 (FM dimer) to $\pi$ (AFM dimer)). The two adatoms are on a 2D superconductor with a square-shaped FC, in three configurations  along the (100), (110) and (210) directions (i.e. with inter-atomic distances 2$a$, 2$\sqrt(2)a$ and $\sqrt(5)a$, respectively). (b)-(e) Evolution of the YSR spectral function with the dimer's separation along the (100) direction for different spin  angles.}
	\label{Fig3}
\end{figure}

We apply our continuum model to obtain the spectral evolution of YSR states as a function of the relative angle between impurities, similarly as in Ref. \onlinecite{Kezilebieke2018} where the calculation was done using a tight-binding lattice. 
Fig. \ref{Fig3} (a) shows the angular dependence of the energy splitting of hybridized YSR states for the three Mn dimers of the experiment shown in Fig.~\ref{Fig2} d).   As mentioned above, the  YSR splitting of a (2$a$,0) dimer is expected to be the largest for parallel spins but quickly reduces with the relative angle, vanishing for AF spins. Figures \ref{Fig3} (b)-(e) show the evolution of the YSR states with dimer separation along the (100) direction for different relative angles of the impurity spins. The oscillation amplitude of even and odd states decreases when they are non-collinear and merge into a single peak for AF spins. These results suggest that Mn dimers build along (100) direction are  antiferromagnetically aligned in contrast to the other dimers explored, whose YSR splitting is consistent with a close to ferromagnetic alignment of their relative spins. We suggest that, in addition to inducing a larger YSR wavefunction hybridization along the (100) direction, the anisotropic FC also leads to stronger exchange interaction between dimers \cite{Yao2014a}, which at such close distances promotes their mutual AF alignment.

\section{Conclusions}

 In conclusion, we present an analytical method to compute the GFs and spectrum of a two  dimensional  superconductor with an arbitrary  FC   in the presence of magnetic impurities.  We apply  the method to FCs with the shape of a regular polygon.  We found that the spatial dependence of the YSR subgap states reflects the symmetry of the FC, and that the characteristic decay length of such states strongly depends on the spatial direction.  Namely, Fermi surface nesting in low symmetry cases lead to a focusing effect of the YSR spectrum. We contrast our model  with a tight-binding model and STM measurements on materials with different FC shapes and find good agreement.  
We also present STS measurements of Mn dimers on top of a superconductor with a square FC and, by  comparing them with our theoretical  results, we demonstrate the  applicability of the approach.

\begin{acknowledgments}
We acknowledge financial support from Spanish AEI through projects PID2019-107338RB-C61, PID2020-117671GB-I00 and 
PID2020-114252GB-I00 (SPIRIT), and through the Maria de Maeztu Units of Excellence Program (MDM-2016-0618) and (CEX2018-000805-M). Jon Ortuzar acknowledges funding provided by BERC Materials Physics Center through its doctorate program and the PRE\_2021\_1\_0350 scholarship from the Basque Government. 
\end{acknowledgments}

\onecolumngrid
\appendix

\section{Integration of several FCs.}\label{app:A}

In this appendix we present the  result of  the  integral Eq. (\eqref{fouriertrans})  for a M-sided regular polygon. We then use the solution to calculate the integral for the square-shaped contour and the $M\to \infty$ limit which corresponds to the circular case.

We define the  basis vectors $(\hat{\textbf{u}}_x,\hat{\textbf{u}}_y)\rightarrow(\hat{\textbf{n}},\hat{\textbf{w}})$, where $\hat{\textbf{n}}$ is a vector normal to a polygon's side and $\hat{\textbf{w}}$ its perpendicular vector. The vertices of the polygon are at the points 
\begin{equation}
    V_i: (\cos(\frac{2i-1}{M}\pi),\sin(\frac{2i-1}{M}\pi))\; .
\end{equation}

It is also convenient to  define   $p_F'=p_F\cos(\frac{\pi}{M})$, where $p_F$ if the Fermi momentum of the inscribing circumference, i.e, the momentum in the vertices of the polygon.  The integral along  one of the polygon sides can be written as
\begin{equation}
    \dfrac{1}{(2\pi)^2}\int \dfrac{d\xi}{p_n}       G_0(\xi)e^{-i(\frac{m\xi}{p_F'}+p_F')n^k}\int_{-p_u^k\tan\frac{\pi}{M}}^{p_n^k\tan\frac{\pi}{M}}e^{-iqw^k}dq=\dfrac{1}{(2\pi)^2}\int_{n^k/w^k-\tan\frac{\pi}{M}}^{u^k/w^k-\tan\frac{\pi}{M}}d\chi \int d\xi G_0(\xi)e^{-i(\frac{m\xi}{p_F'}+p_F')\chi w^k}\; ,
\end{equation}
where we defined, $n^k=x\cos(\frac{2\pi k}{M})+y\sin(\frac{2\pi k}{M})$, $p_n^k=p_x\cos(\frac{2\pi k}{M})+p_y\sin(\frac{2\pi k}{M})$, $w^k=-x\sin(\frac{2\pi k}{M})+y\cos(\frac{2\pi k}{M})$ and $p_w^k=-p_x\sin(\frac{2\pi k}{M})+p_y\cos(\frac{2\pi k}{M})$. Thus the  GF reads
\begin{equation}\label{cg0}
    G_0(\textbf{x})=\sum_{k=1}^{M}\dfrac{1}{(2\pi)^2}\int_{n^k/w^k-\tan\frac{\pi}{M}}^{n^k/w^k+\tan\frac{\pi}{M}}d\chi \int d\xi  G_0(\xi)e^{-i(\frac{m\xi}{p_F'}+p_F')\chi w_k}\; .
\end{equation}

Defining $\theta_p=\frac{2\pi k}{M}$, in the limit $M\rightarrow\infty$ we get that
\begin{equation}
    G_0(\textbf{x})=\sum_{k=1}^{N}\dfrac{1}{(2\pi)^2}\int_{n(\theta_p)/w(\theta_p)-\frac{d\theta}{2}}^{n(\theta_p)/w(\theta_p)+\frac{d\theta}{2}}d\chi \int d\xi  G_0(\epsilon,\xi)e^{-i(\frac{m\xi}{p_F}+p_F)\chi w(\theta_p)}=\int_{0}^{2\pi}\dfrac{d\theta}{(2\pi)^2}\int d\xi  G_0(\epsilon,\xi)e^{-i(\frac{m\xi}{p_F}+p_F)u(\theta_p)}\
\end{equation}

Note that $u(\theta_p)=x\cos(\theta_p)+y\sin(\theta_p)=r[\cos(\theta_p)\cos(\theta_r)+\sin(\theta_p)\sin(\theta_r)]=r\cos(\theta_p-\theta_r)$, so;
\begin{equation}\label{circ}
    G_0(\textbf{x})=\int_{0}^{2\pi}\dfrac{d\theta_p}{(2\pi)^2}\int d\xi  G_0(\epsilon,\xi)e^{-i(\frac{m\xi}{p_F}+p_F)r\cos(\theta_p-\theta_r)}\
\end{equation}

Integral \eqref{circ} corresponds to the circular FC, which can be analytically solved.
\begin{equation}
\begin{split}
    G_0(\textbf{x})&=\dfrac{m}{(2\pi)^2}\int_{0}^{2\pi}d\theta_pe^{ip_Fr\cos(\theta_p-\theta_r)}\int_{-\infty}^{\infty}d\xi \dfrac{\epsilon\tau_0\sigma_0+\Delta\tau_1\sigma_0+\xi\tau_3\sigma_0}{\xi^2+\omega^2}e^{i\frac{m}{p_F}\xi r\cos(\theta_p-\theta_r)}\\
    &=\dfrac{m}{4\pi}\int_{0}^{2\pi}d\theta_p(\rho_{BCS}(\omega)+i\text{sign}(\cos(\theta_p-\theta_r))\tau_3\sigma_0)e^{-\frac{m}{p_F}\omega r|\cos(\theta_p-\theta_r)|+ip_Fr\cos(\theta_p-\theta_r)}\; ,
\end{split}
\end{equation}
where $\rho_{BCS}(\omega)=\frac{\epsilon\tau_0\sigma_0+\Delta\tau_1\sigma_0}{\omega}$, with $\omega=\sqrt{\Delta^2-\epsilon^2}$. The integral is easier to evaluate after making the change $\theta'=\theta_p-\theta_r$. Noticing that 
\begin{equation}
    \int_{-\frac{\pi}{2}}^{\frac{\pi}{2}}d\theta e^{i u \cos(\theta)}=\pi\left( J_0(u)+iH_0(u) \right)\; ,
\end{equation}
where $J_0(x)$ is the zero order Bessel function of the first type and $H_0(x)$ is the zero order Struve function, we finally obtain:
\begin{equation}
    G_0(\textbf{r})=\pi N_0\left\{ \rho_{BCS}(\omega)\left[ \text{Re}\{J_0(u)+iH_0(u)\} \right]+ i\tau_3\sigma_0\left[ \text{Im}\{J_0(u)+iH_0(u)\} \right]\right\}\; ,
\end{equation}
where $u=(p_F+ i\frac{m}{p_F}\omega) r=(p_F+i\xi_s^{-1})r$ and $N_0$ is the normal metal DoS. In the limit $r\rightarrow 0$ this expression reduces to the  BCS Green's function. 
In the asymptotic limit $p_F r\gg1$ we obtain\cite{PhysRevB.91.064505}:
\begin{equation}\label{esph_asymp}
    G_0(E,\textbf{x})=\pi N_0i\tau_3\sigma_0\left[ \sqrt{\dfrac{2}{\pi p_F r}}\sin\left(p_Fr -\dfrac{\pi}{4}\right)e^{-r/\xi_s}+\dfrac{2}{\pi p_F r} \right]+\pi N_0\rho_{BCS}(\omega) \sqrt{\dfrac{2}{\pi p_F r}}\cos\left(p_Fr -\dfrac{\pi}{4}\right)e^{-r/\xi_s}
\end{equation}

We now calculate the integral for a square-shaped FC. Taking $M=4$ in Eq. \eqref{cg0} one can straightforwardly check that:
\begin{equation}
\label{G0-square}
\begin{split}
   G_0(\textbf{x})&=\dfrac{m}{2\pi y}\int_{x-y}^{x+y} dv (\rho_{BCS}(\omega)\cos(p_Fv)-\sin(p_F|v|)\tau_3\sigma_0)e^{-\frac{m}{p_F}\omega |v|}\\
   &+\dfrac{m}{2\pi x}\int_{y-x}^{x+y} dv (\rho_{BCS}(\omega)\cos(p_Fv)-\sin(p_F|v|)\tau_3\sigma_0)e^{-\frac{m}{p_F}\omega |v|}
\end{split}
\end{equation}
All these integrals are analytically solvable. 
We focus here on the region $x+y>0$ and $x-y>0$ (due to the symmetry of the system results for the other  regions are obtained similarly). In leading order in $m\omega/p_F^2\ll1$ we obtain
\begin{equation}
\begin{split}
   G_0(\textbf{x})=&\pi\dfrac{N_0}{4} \dfrac{\rho_{BCS}(\omega)}{p_F^2}\left\{e^{-\frac{m}{p_F}\omega(x+y)}\left(\dfrac{1}{x}+\dfrac{1}{y}\right)p_F\sin(p_F(x+y)) +e^{-\frac{m}{p_F}\omega(x-y)}\left(\dfrac{1}{x}-\dfrac{1}{y}\right)p_F\sin(p_F(x-y))\right\}\\
   &-\pi\dfrac{N_0}{4}\dfrac{\tau_3\sigma_0}{p_F^2}\left\{e^{-\frac{m}{p_F}\omega(x+y)}\left(\dfrac{1}{x}+\dfrac{1}{y}\right)p_F\cos(p_F(x+y))+e^{-\frac{m}{p_F}\omega(x-y)}\left(\dfrac{1}{x}-\dfrac{1}{y}\right)p_F\cos(p_F(x-y)) -\dfrac{2p_F}{x} \right\}\\
   &+\mathcal{O}\left(\dfrac{m\omega}{p_F}\right)\; .
\end{split}
\end{equation}

In the main text we discuss the GF over the  lines $y=0$ and $x=y$: \begin{equation}\label{y=0}
\begin{split}
   G_0(x,y=0)=&\pi N_0\rho_{BCS}(\omega)\left\{e^{-|x|/\xi_s}\left[\dfrac{1}{p_F x}\sin(p_F x) + \cos(p_F x) \right] \right\}\\
   &-\pi N_0\tau_3\sigma_0\left\{e^{-|x|/\xi_s}\left[\dfrac{1}{p_Fx}\cos(p_Fx) -\sin(p_Fx) \right] -\dfrac{1}{p_Fx}\right\}
\end{split}
\end{equation}
\begin{equation}\label{x=y}
   G_0(x=y)=\dfrac{m}{\pi }\rho_{BCS}(\omega)\left\{e^{-2x/\xi_s}\dfrac{1}{p_Fx}\sin(2p_Fx) \right\}-\dfrac{m}{\pi}\tau_3\sigma_0\left\{e^{-2x/\xi_s}\dfrac{1}{p_Fx}\cos(2p_Fx)-\dfrac{1}{p_Fx} \right\}
\end{equation}

\twocolumngrid
\bibliographystyle{apsrev4-1}

%

\end{document}